# Music Genre Classification: A Comparative Analysis of CNN and XGBoost Approaches with Mel-frequency cepstral coefficients and Mel Spectrograms

Yigang Meng[1, a)]

*Department of Statistics, University of California, Davis, 95616, United States*
*Corresponding author: mgmeng@ucdavis.edu*

**Abstract.** In recent years, various well-designed algorithms have empowered music platforms to provide content based on one's preferences. Music genres are defined through various aspects, including acoustic features and cultural considerations. Music genre classification works well with content-based filtering, which recommends content based on music similarity to users. Given a considerable dataset, one premise is automatic annotation using machine learning or deep learning methods that can effectively classify audio files. The effectiveness of systems largely depends on feature and model selection, as different architectures and features can facilitate each other and yield different results. In this study, we conduct a comparative study investigating the performances of three models: a proposed convolutional neural network (CNN), the VGG16 with fully connected layers (FC), and an eXtreme Gradient Boosting (XGBoost) approach on different features: 30-second Mel spectrogram and 3-second Mel-frequency cepstral coefficients (MFCCs). The results show that the MFCC XGBoost model outperformed the others. Furthermore, applying data segmentation in the data preprocessing phase can significantly enhance the performance of the CNNs.

## INTRODUCTION

The classification of musical genres has long been discussed and implemented in scientific and industrial fields. It is a subset of Automatic Music Retrieval (AMR). Two main techniques build an effective recommendation system: collaborative filtering and content-based filtering. Collaborative filtering provides recommendations based on user proximity [1]. Previous research has identified two main challenges arising from the inherent nature of sparse matrices and algorithm design. The first challenge is the cold start problem, which introduces new users to the system who need more information to generate predictions [2]. Popularity bias can occur within algorithms, causing the system to frequently prioritize recommending popular items and ignore lesser-known options [3]. One way to address this problem is by employing well-designed algorithms with content-based filtering. This method extracts content information based on music annotations or classifications and does not require user information.

### Music Genre Classification Methods

Various methods of music genre classification have been extensively researched, including deep learning and traditional methods. C. Liu et al. have significantly contributed to a state-of-the-art CNN that can analyze time-frequency information at multiple scales, achieving 93.9% accuracy on the GTZAN dataset. Moreover, Elbir and Aydin proposed a hybrid architecture that combines CNN with support vector machines (SVM), achieving an accuracy of 97.6% [4].

### Purpose

The implemented methods with features are shown in Table 1. The study focuses on the effectiveness comparison between three methods in the aspects of:

- Assessing the effectiveness of each model with different extracted features.
- Comparing the performance of two CNN-based models and an XGBoost model.
- Referring to former studies to analyze the found effectiveness gaps among models and features

**Table 1.** Implemented models and features

| Model | Feature |
|---|---|
| CNN | 3s MFCC |
| VGG16 + Fully connected layers (FC) | Mel Spectrogram |
| XGBoost | 3s MFCC |

## METHODS

### GTZAN Dataset

The GTZAN dataset has 1000 audio files, each having 30 seconds duration, and it is evenly classified into ten classes:
1. Blues
2. Classical
3. Country
4. Disco
5. Hip-hop
6. Jazz
7. Metal
8. Pop
9. Reggae
10. Rock

In preprocessing, a corrupted file was identified within the jazz genre category. For the CNN and VGG16 with fully connected layers models, the GTZAN data is split into training, testing, and validation datasets with 80%, 16%, and 4% percentages.

### Features Extraction

In music genre classification and audio recognition, two features are primarily used and discussed: MFCCs and mel Spectrograms. They are used in the classification methods.

*Mel Spectrogram*

Mel Spectrograms are two-dimensional representations of the audio. They are in the time-frequency domain; thus, to obtain mel spectrograms, a generic pipeline should be:
1. Framing and windowing the signals,
2. Applying the Fast Fourier Transform (FFT), and
3. Projecting onto the mel scale and taking the logarithms of the frequencies

The mel spectrograms, whose size is 432 by 288, are provided in the GTZAN dataset, each 30 seconds long. Several mel spectrograms are shown in Figure 1. Modifying the image into 3D arrays is necessary when proceeding to the model part.

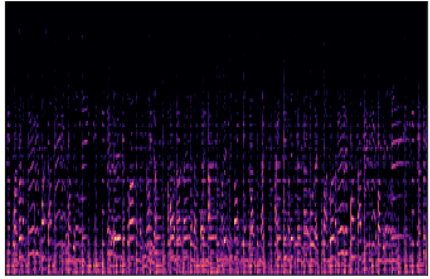 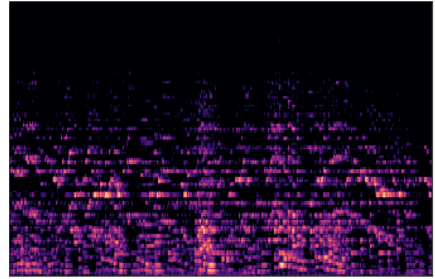

(a) Blues  (b) Classical

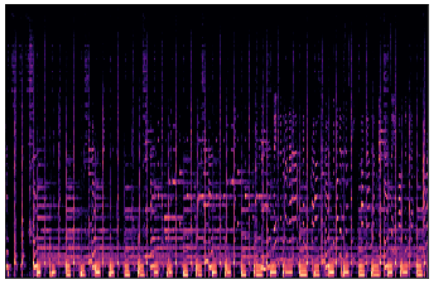 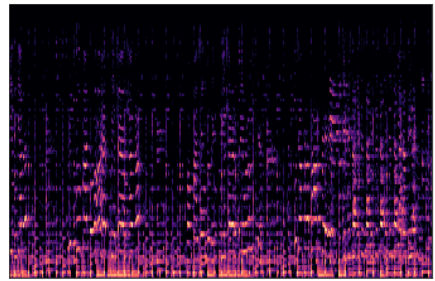

(c) Disco  (d) Reggae

**FIGURE 1.** Mel Spectrograms

*MFCCs*

On the other hand, MFCCs represent a set of features extracted from the mel spectrograms using discrete cosine transform (DCT). For the extraction of MFCCs, the librosa library was utilized in this study. The GTZAN audio files were used as input data and underwent the framing process, resulting in 3-second-long frames. The extraction parameters for MFCCs are:
1. Number of MFCCs (n_mfcc): 13
2. FFT Window Size (n_fft): 2048
3. Number of segments per track: 10
4. Hop length: 512

The dataset was expanded to ten times its original size through signal framing techniques. Figure 2 shows the distribution of the MFCCs. The extracted MFCCs have the size of (9986, 130, 13), which indicates that there are 9,986 segments, 130 MFCC vectors in each segment, and 13 MFCCs are computed. Additionally, the dataset is balanced across all genres.

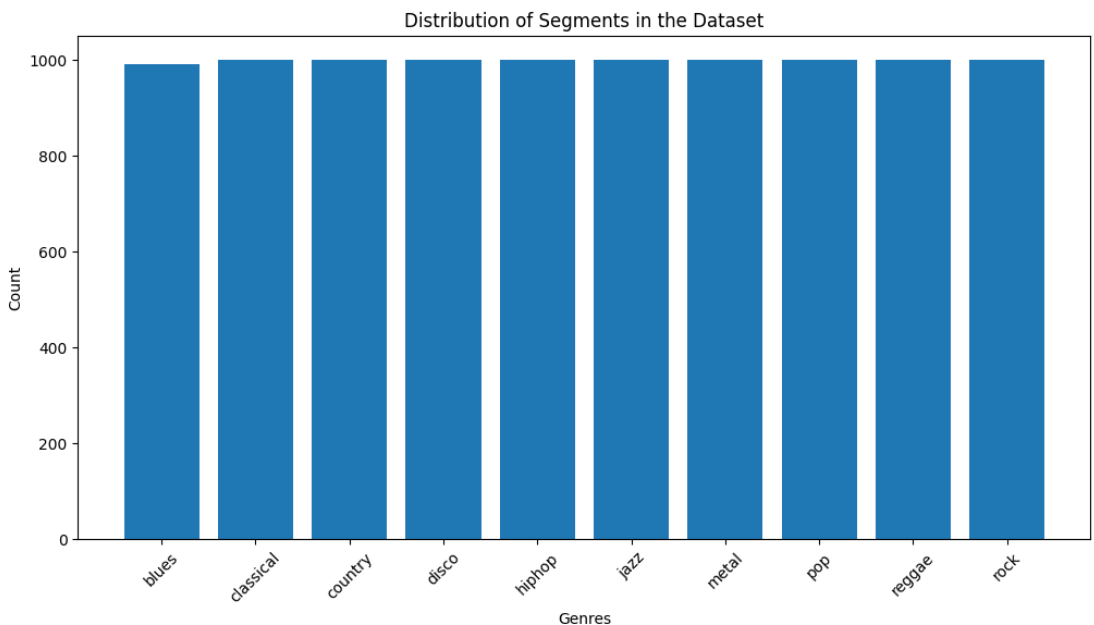

**FIGURE 2.** Distribution of Segments Across Genres in the Dataset

## Models

Three models are implemented in the study: CNN, VGG16 with fully connected layers, and XGBoost. The MFCC CNN are selected and implemented as the baseline model for the study. While few comparative studies are found investigating the performances of MFCCs and mel spectrograms, previous research has suggested that when of the same length, the mel spectrogram can be more effective and accurate when using the CNN in raga classification than the MFCCs [5].

*CNN with 3-second MFCCs*

Table 3 showcases the baseline model utilizing TensorFlow and Keras libraries, which have three convolutional layers, each with 32 kernels. Rectified Linear Unit (ReLU) and Softmax are the activation functions throughout the layers:

$$ReLU(x) = \max(0, x) \tag{1}$$

$$Softmax(x_i) = \frac{e^{x_i}}{\Sigma_{j=1}^{n} e^{x_j}}, \tag{2}$$

where the $i$-th element in a vector $x$ of $n$ real number.

After the convolutional layers, Max-pooling layers are employed to decrease the spatial size of the convolutional layers' outputs. Batch normalization layers follow to introduce stability into the model. The utilization of a dropout layer with a 0.3 rate mitigates overfitting. The fully connected layers (FC) initially flatten matrices into a long vector, followed by two dense layers. The first dense layer uses ReLU activations, and the final Softmax activation generates the probability distribution of genres.

**Table 2.** Architecture of the implemented CNN

| Layer | Output Shape | Param # |
|---|---|---|
| Conv2D | (None, 128, 11, 32) | 320 |
| MaxPooling2D | (None, 64, 6, 32) | 0 |
| BatchNormalization | (None, 64, 6, 32) | 128 |
| Conv2D | (None, 62, 4, 32) | 9248 |
| MaxPooling2D | (None, 31, 2, 32) | 0 |
| BatchNormalization | (None, 31, 2, 32) | 128 |
| Conv2D | (None, 30, 1, 32) | 4128 |
| MaxPooling2D | (None, 15, 1, 32) | 0 |
| BatchNormalization | (None, 15, 1, 32) | 128 |
| Dropout | (None, 15, 1, 32) | 0 |
| Flatten | (None, 480) | 0 |
| Dense | (None, 64) | 30784 |
| Dense | (None, 10) | 650 |

*VGG16-based CNN with Mel Spectrograms*

Table 3 shows the architecture of the VGG16 CNN with the fully connected layers. Notice that the model has been set to non-trainable to preserve the features and reduce computational time. The only trainable part is the proposed fully connected layers. It comes with a dense layer with 128 neurons, a rate of 0.5 dropout layer, and a dense layer with Softmax activation. A batch normalization layer is after the dropout layer to promote stability.

**Table 3.** Architecture of the VGG16 + FC

| Layer | Output Shape | Param # |
|---|---|---|
| BatchNormalization | (None, 288, 432, 3) | 12 |
| Conv2D | (None, 288, 432, 32) | 896 |
| MaxPooling2D | (None, 143, 215, 32) | 0 |
| Conv2D | (None, 141, 213, 32) | 9248 |
| MaxPooling2D | (None, 70, 106, 32) | 0 |
| Conv2D | (None, 68, 104, 32) | 9248 |
| MaxPooling2D | (None, 34, 52, 32) | 0 |
| Conv2D | (None, 32, 50, 32) | 9248 |
| MaxPooling2D | (None, 16, 25, 32) | 0 |
| Conv2D | (None, 14, 23, 64) | 18496 |
| MaxPooling2D | (None, 7, 11, 64) | 0 |
| Flatten | (None, 4928) | 0 |
| Dense | (None, 128) | 630912 |
| Dropout | (None, 128) | 0 |
| BatchNormalization | (None, 128) | 512 |
| Dense | (None, 10) | 1290 |

*XGBoost with 3-second MFCCs*

eXtreme Gradient Boosting (XGBoost) classifier is implemented in the study. XGBoost is a decision-tree-based algorithm that proves effective on tabular datasets [6]. It inherits the idea of the gradient boosting technique. Simply put, it starts with an imperfect model $F_0$ that predicts the true value $y_i$ by adding $h_m(x)$, where $m$ is the stage, and $1 \leq m \leq M$. Instead of directly optimizing $h_m$, it is a good practice to leverage the idea of gradients:

$$L = E\left((y_i - F(x_i))^2\right) \tag{3}$$

$$-\frac{\partial L}{\partial F(x_i)} = \frac{2}{n} h_m(x_i) \tag{4}$$

$$F_m(x) = F_{m-1}(x) - \gamma \sum_{i=1}^{n} \nabla_{F_{m-1}} L(y_i, F_{m-1}(x_i)), \tag{5}$$

where the loss function employed is mean squared error. Hence, the learning rate $\gamma_m$ is being determined by the equation:

$$\gamma_m = \arg\min_{\gamma} \sum_{i=1}^{n} L(y_i, F_{m-1}(x_i)) - \gamma \nabla F_{m-1} L(y_i, F_{m-1}(x_i)) \quad (6)$$

The XGBoost model has several estimators of 1,000, while others remain default.

# RESULTS

Table 4 showcases that the XGBoost method with 3s MFCC achieves the highest performance among the three. The testing AUC is 0.9682, and the testing accuracy is 97%. The CNN model with 3s MFCC follows with 0.9930 and 89% for testing AUC and accuracy, respectively. The VGG16 achieves the lowest scores with 0.9516 and 71% for the AUC and accuracy.

**Table 4.** Comparison of AUC and accuracy for models

| Model | AUC | Testing Accuracy |
|---|---|---|
| CNN with 3s MFCC | 0.9947 | 91% |
| VGG16 + FC with 30s Mel spectrogram | 0.9516 | 71% |
| XGBoost with 3s MFCC | 0.9682 | 97% |

The proportional confusion matrices are provided in Figure 3, given the fact that two different strategies of segmenting are used throughout the study. For the overall performance, the XGBoost method outperforms others as the matrix closely resembles an identity matrix. The confusion matrix of the CNN model with 3-second MFCC is overall decent, while the classical and pop segments are the most mislabeled. The VGG16 with 30s mel spectrograms is significantly outperformed by others, where rock music has the lowest precision. Pathania and Bhardwaj have indicated that the VGG16 transfer learning model has an accuracy of 63% on 10-second mel spectrograms [7]. Though Bhardwaj used a different dataset with a larger size, the VGG16 model outperformed other feature engineering-based models like logistic regression and support vector machines. VGG16 also proved effective and outperformed CNN on the GTZAN mel spectrogram with the same length [8]. Hence, it's plausible to infer that data segmentation can promote accuracy for CNN models in music genre classification to some extent.

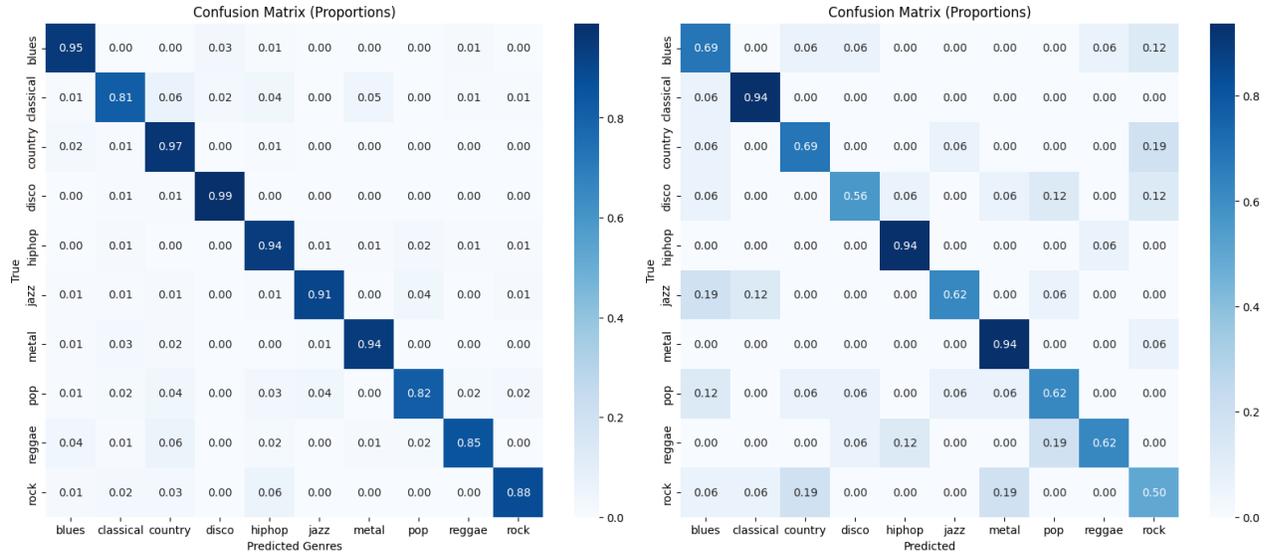

(a) CNN proportional confusion matrix    (b) VGG16 + FC proportional confusion matrix

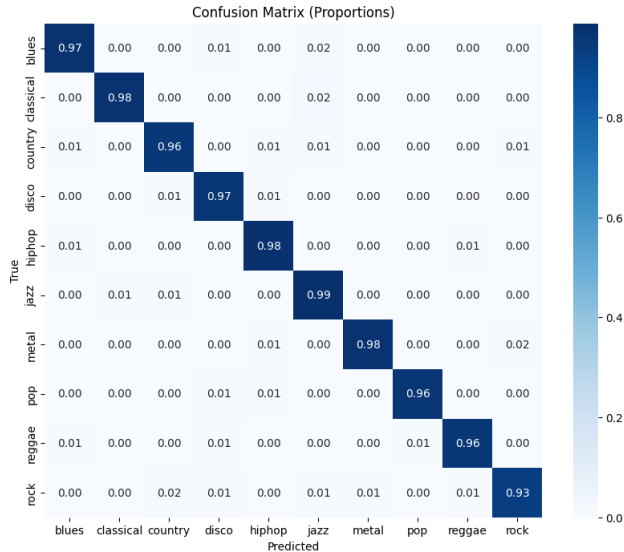

(c) XGBoost proportional confusion matrix

**FIGURE 3.** Proportional confusion matrices for three models

Furthermore, the loss and accuracy graphs for deep learning methods along the epochs are shown in Figure 4. Early stops are implemented in each to prevent over-fitting. The convergence for each is similarly fast, while the loss and accuracy graph for VGG16 with 30-second mel spectrograms fluctuates more than the CNN with 3-second MFCC. Moreover, the GTZAN dataset proves somewhat ineffective and limited [9], which may further explain why the CNN models fall behind as CNNs typically require large annotated datasets [10].

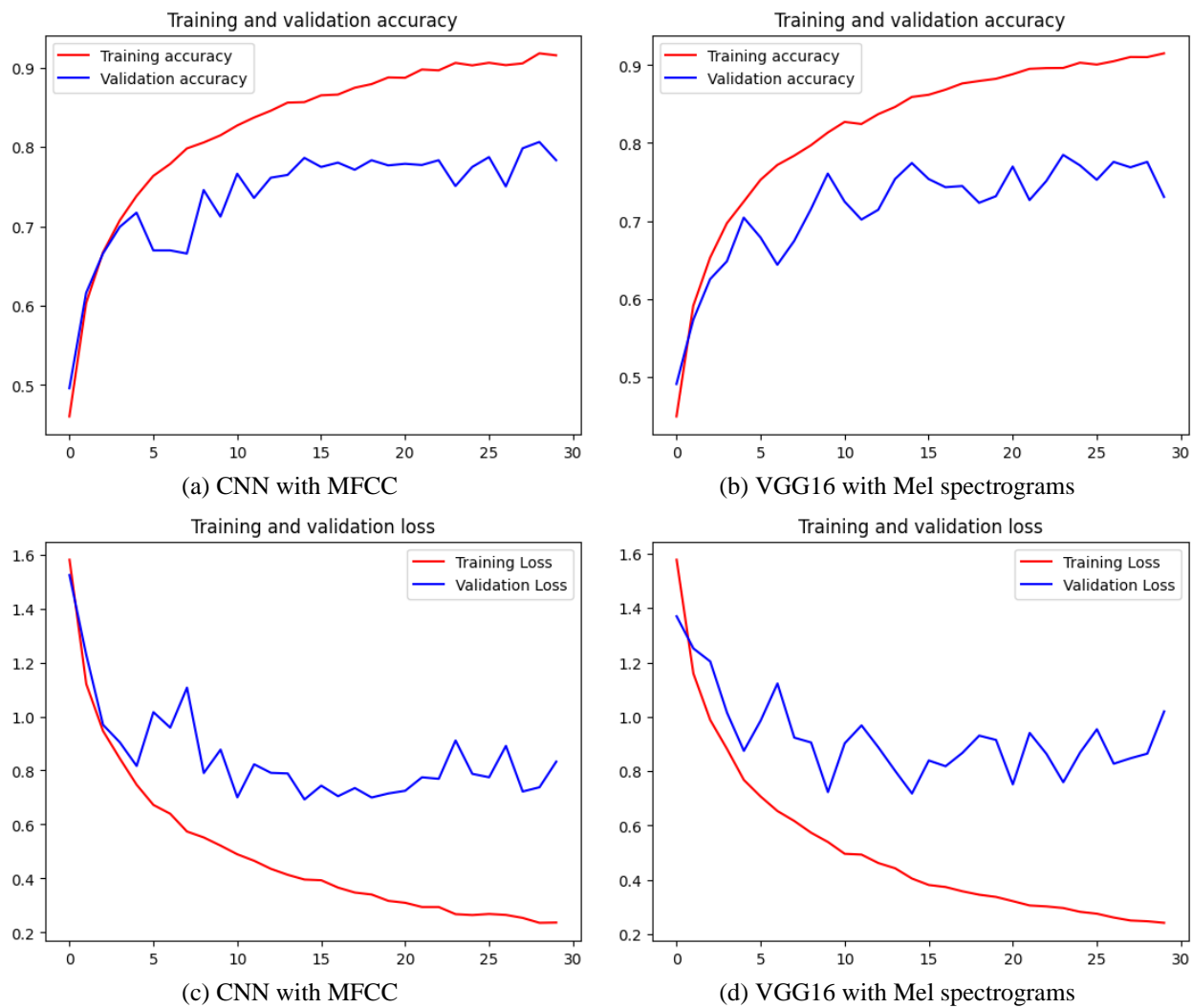

**FIGURE 4.** Loss and accuracy graphs of CNN and VGG16 + FC

The XGBoost achieves the best results where the tabular data is critical. The data extraction transformed the 2D representation into table-like data with numerical numbers. Ziv and Armo has shown that the proposed XGBoost transcended other deep models on the tabular datasets that didn't originate from their papers [6]. Unexpectedly, XGBoost is mainly considered a baseline, and what it achieved may be more than people assume, including this study.

## CONCLUSION

This study implements the CNN with 3s MFCC, VGG16 with mel spectrogram, and XGBoost with MFCC. From the results, the XGBoost outperforms others as MFCC is tabular. Therefore, XGBoost is a competitive candidate when providing a tabular dataset. In addition, based on the resultant comparison between CNN and VGG16 and the former studies, a conclusion can be drawn that data segmentation can promote model accuracy in music genre classification tasks. However, it is insufficient as it is not comprehensive and does not contain enough segmenting strategies.

Hence, further work is needed to build a more uniform, comprehensive dataset. In addition, given that music can have various lengths and the genre is developing much faster nowadays, model design and evaluation are expected to be more mature to manage these factors systematically.